\documentclass{emulateapj}

\shorttitle{Double QSOs from SDSS}
\shortauthors{Rosario et al.}

\newcommand{\kms}{km s$^{-1}$}
\newcommand{\hb}{H$\beta$}
\newcommand{\othree}{[\ion{O}{3}]$\lambda 5007$}

\begin{document}

\title{Adaptive Optics Imaging of QSOs with Double-Peaked Narrow Lines: Are they Dual AGNs?}

\author{D.J. Rosario\altaffilmark{1}, R.C. McGurk\altaffilmark{2}, C.E. Max\altaffilmark{2}, 
G.A. Shields\altaffilmark{3}, K.L. Smith\altaffilmark{3}, \& S.M. Ammons\altaffilmark{4}}

\altaffiltext{1}{Max-Planck-Institute for Extraterrestrial Physics,
Garching, 85748; rosario@mpe.mpg.de}

\altaffiltext{2}{Astronomy Department and UCO-Lick Observatory, University of California, Santa Cruz, CA 95064; mcgurk@ucsc.edu, max@ucolick.org}

\altaffiltext{3}{Astronomy Department, University of Texas, Austin,
TX 78712; shieldsga@mail.utexas.edu, krista@mail.utexas.edu}

\altaffiltext{4}{Steward Observatory, University of Arizona, Tucson,
AZ  85721; ammons@as.arizona.edu}

\begin{abstract}

Active galaxies hosting two accreting and merging super-massive black holes (SMBHs) -- dual Active Galactic Nuclei (AGN) -- are predicted by many current and popular models of black hole-galaxy co-evolution. We present here the results of a program that has identified a set of probable dual AGN candidates based on near Infra-red (NIR) Laser Guide-Star Adaptive Optics (LGS AO) imaging with the Keck II telescope. These candidates are selected from a complete sample of radio-quiet Quasi-stellar Objects (QSOs) drawn from the Sloan Digital Sky Survey (SDSS), which show double-peaked narrow AGN emission lines. Of the twelve AGNs imaged, we find six with double galaxy structure, of which four are in galaxy mergers. We measure the ionization of the two velocity components in the narrow AGN lines to test the hypothesis that both velocity components come from an active nucleus. The combination of a well-defined parent sample and high-quality imaging allows us to place constraints on the fraction of SDSS QSOs that host dual accreting black holes separated on kiloparsec (kpc) scales: $\sim 0.3-0.65$\%. We derive from this fraction the time spent in a QSO phase during a typical merger and find a value that is much lower than estimates that arise from QSO space densities and galaxy merger statistics. We discuss possible reasons for this difference. Finally, we compare the SMBH mass distributions of single and dual AGN and find little difference between the two within the limited statistics of our program, hinting that most SMBH growth happens in the later stages of a merger process. 

\end{abstract}

{\it Facilities:} \facility{Keck:II (laser guide star adaptive optics, NIRC2)}

\section{Introduction}

Active Galactic Nuclei (AGN) fueled by major galaxy mergers are predicted from models of black hole-galaxy co-evolution, in which black holes and their host galaxies grow together as an outcome of hierarchical structure formation in the universe \citep[e.g.][]{ferrarese00, gebhardt00}. In this scenario, powerful tidal forces funnel material into the centers of merging gas-rich galaxies, driving the growth of super-massive black holes (SMBHs) at near-Eddington rates. One prediction of this type of model is that prior to the final merger event, some ongoing gas-rich mergers should host dual AGNs separated by spatial scales comparable to the sizes of the galaxies.

Two sets of factors determine the fraction of merging galaxies that harbor dual AGN. One is the causal link between galaxy mergers and high accretion rate phases of SMBH growth. Studies show that most low-luminosity AGNs (i.e., Seyferts) are not undergoing significant mergers \citep{cisternas10}, though they may be frequently associated with galaxy-scale tidal interactions revealed in HI \citep{kuo08}. However, powerful AGNs (i.e. QSOs) are believed to be strongly correlated with major merger events \citep{hopkins06, hopkins09}. Such studies have been borne out by associations between statistically determined quantities such as the quasar luminosity function, galaxy and merger mass functions, and quasar clustering \citep{hopkins08}.

The second key factor is the frequency at which both black holes in a merger accrete gas and shine together as AGN, while still separated on scales of a few kpc. This fraction is not well constrained. Most computational models do not address pre-coalescence fueling very accurately, since it depends on several parameters that lie below the resolution of typical simulations. If a substantial fraction of QSOs show dual active nuclei and signatures of strong activity in the early or intermediate stages of a galaxy merger, a significant portion of the growth of the SMBH will occur before final coalescence. On the other hand, if a great majority of QSOs are seen as single AGN, the SMBHs in these systems have either in-spiralled to parsec separations or have already coalesced before substantial activity begins. In such a case, most black hole growth is expected to occur in the late part of a merger. Knowledge of the phase during which most of the accretion and QSO activity takes place has important consequences for the form and scatter of SMBH scaling laws, because black hole growth in a gas-rich galaxy merger runs in parallel with another important avenue of gas depletion: star-formation. If a large fraction of SMBH growth occurs early in the merger, the eventual relationship between the mass of the black hole and the stellar content of the merger remnant will be influenced.

The incidence of dual AGNs in an ongoing merger therefore provides insight into the physical processes taking place when the two merging SMBHs are separated on galaxy scales or less. In the past few years, studies have searched large extra-galactic spectroscopic datasets to arrive at statistically useful samples of  double QSO or AGN candidates \citep{hennawi06,  myers07, myers08, hennawi10} with separations on the sky up to 650 kpc. More recent studies identified pairs of QSOs separated on scales of tens of kpc \citep{foreman09} to a few kpc \citep{junkkarinen01, comerford09, liu09, smith10, wang09}. From such studies, it is clear that the fraction of candidate dual AGNs among QSOs is small, even after accounting for the short duty cycle of phases with high accretion rates and the relative time-scales during which both nuclei are visible as AGNs \citep{smith10}. Note that the identification of double AGN separated by a few kpc from spectroscopic surveys gives an upper limit to the dual AGN fraction, since such studies select for kinematic substructure in narrow emission lines which may also arise from AGN jets or outflows.

In this paper we approach the issue by looking for the ``smoking gun": individual galaxies which contain dual AGNs and whose morphologies are typical of merging systems. 
To place constraints on the fraction of QSOs which show double structure on scales of 10 kpc or less, we have undertaken a program of near-IR (NIR) imaging of candidate radio-quiet double QSOs using the Keck Laser-Guide Star Adaptive Optics (LGS AO) system \citep{wizinowich06}. In this Paper, from our 12 targets we report six dual AGN candidates, four in close mergers and two in double systems, with separations ranging from 3 to 12 kpc. We measure luminosities of the AGN and host galaxies, the separations of the merging components, as well as black hole masses of the brighter QSO in each pair.  Since we start with a complete parent sample of QSOs, we are able to refine the fraction  which show double structure on scales larger than 0.45-1.6 kpc (over our redshift range), and we discuss the frequency of such closely separated merging AGN in the SDSS QSO population at large. We compare our results with a recent study by \citet{fu10} which images a larger sample of Type I and Type II AGN to a shallower depth. Many of the double systems presented in this paper can also be found in \citet{fu10}. We adopt a Concordance Cosmology with $H_{0} = 70$ km/s/Mpc and $\Omega_{\Lambda} = 0.7$. 

\begin{table*}[t]
\caption{Sample of QSOs with NIRC-2 LGS AO imaging}
\begin{center}
\begin{tabular}{cccccccc}
\hline
SDSS ID & z & Double? & Sep. (") & Sep. (kpc) & m$_{\textrm{H}}$ (QSO) & m$_{\textrm{H}}$ (host) & m$_{\textrm{H}}$ (companion) \\
\hline
\hline
J153231.80+420342.7   & 0.210   & n & N/A & N/A    & 16.81  &  16.54 &  ---  \\
J091649.41+000031.5   & 0.222   & n & N/A & N/A    & --- & --- & --- \\
J161027.41+130806.8   & 0.229   & y & 2.35 & 8.55   & 16.87  &  15.77  &  19.98 \\
J081542.53+063522.9   & 0.244   & n & N/A & N/A    &  16.46  &  16.01 & --- \\
J095207.62+255257.2   & 0.339   & y & 1.00 & 4.82   &  17.96  &  16.36  & 17.43 \\
J130724.08+460400.9   & 0.353   & y & 2.37 & 11.64 &  20.47  &  17.62  &  17.86 \\
J020011.52-093126.1    & 0.361   & y & 1.17 & 5.84   &  16.31  &  16.93  &  19.55 \\
J140923.51-012430.5    & 0.405   & n & N/A & N/A    & 16.65  &  19.13 & --- \\
J124859.72-025730.7    & 0.487   & y & 0.53 & 3.15   & 16.58 &   17.22 &   16.82 \\  
J154107.81+203608.8   & 0.508   & y & 2.00 & 12.22 &  17.88  &  18.68  &  19.87 \\
J121911.16+042905.9   & 0.555   & n & N/A & N/A    & 16.01 &   20.10 & --- \\
J072554.42+374436.9   & 0.634   & n & N/A & N/A    & 16.59  &  18.74 & --- \\
\hline
\end{tabular}
\end{center}
\end{table*}

\section{Keck Laser Guide Star AO Imaging Program}

For a well-defined sample of pre-coalescence dual AGN candidates, we drew upon the catalog of \citet{smith10},
which selected SDSS DR7 spectroscopic QSOs that show a double-peaked \othree\ line. Many of them also show similar line-splitting in other bright emission lines, such as [\ion{O}{2}]$\lambda 3727$ and \hb. The full catalog contains 86 Type I AGN. We excluded radio-detected QSOs, which are frequently associated with disturbed emission line regions due to jet interactions \citep{wilson80, whittle92, rosario10}. We also excluded Type II AGN in the SDSS QSO database which are frequently failures in the QSO spectral identification algorithm and hence have rather unconstrained selection criteria \citep{smith10}. To our working sample, we applied LGS AO observability requirements\footnote{An $r<17.5$ tip-tilt (TT) star within an arcminute of the galaxy for a good tip-tilt correction.} which yielded 37 possible LGS AO targets. These were then ranked according to the brightness and nearness of their tip-tilt (TT) stars (which determines the performance of the AO system), the strength of galaxy spectral features in their SDSS spectra (which determines the contrast between the AGN point source and the extended galaxy light), their observed SDSS magnitude (which determines the S/N of the final images) and the quality of the \othree\ line split, following criteria discussed and applied in \citet{smith10}. 

Over a full night on Sep 20 2009 and two half-nights on Jan 4 and Apr 20 2010, a total of 12 QSOs were imaged.  The NIRC2 near-infrared camera (PI: Keith Matthews) on the Keck II telescope was used in the Wide-Camera
mode (FOV: $40$", pixel scale: $0.04$'') with a broadband H filter. After imaging each object, a Point-Spread Function (PSF) reference star pair was imaged with similar TT star properties as those of the science target. This ensured that the PSF reference suffered consistent anisoplanatic effects as the target QSO. Where a good PSF reference pair was not available, a series of short exposures of the TT star was taken instead. 

Exposure times ranged from $20-60$ min depending on the brightness of the target. Large numbers of
dithered sub-exposures were taken to prevent saturation of the sky background, and to overcome flat-field variations across the detector. A similar process was used for the PSF reference pair images, though the sub-exposure 
integration times were much shorter to prevent detector saturation. Sub-exposures were combined in post-processing using a version of the \anchor{http://www.ucolick.org/~jmel/cats_database/nirc2_reduce.tar.gz}{NIRC2 IDL pipeline}, modified to robustly account for the very narrow core of the PSF. This is critical when modeling the QSO host galaxy to estimate the brightness of the central point source.  

\section{ Photometry }

Photometry was performed directly from the NIRC2 images. The NIRC2 detector in our default
operating mode (Multiple-Correlated Double Sampling or MCDS) is known to show a considerable excess of pixels with low counts, believed to be due to the rounding of low valued pixels at readout \footnote{See \anchor{http://www2.keck.hawaii.edu/inst/nirc2/nirc2_news.html}{the NIRC2 News page}}. The excess tends to skew estimates of the local background from modal statistics. We instead fit a gaussian to the local pixel distribution around the object, excluding the pixel values around zero and use the mean of the gaussian as the local background.  A typical absolute difference in the local sky 
estimate from modal statistics vs. the gaussian fit are around 20\%\, but this can be more pronounced in images where the local background is higher. The effect of the low pixel excess on photometry after sky subtraction is negligible. 

All photometry was referred to the 2MASS H-band magnitude system. For objects which had a nearby bright star in the field, the star was used as a photometric calibration standard to the 2MASS system. For the rest, the total photometry of the QSOs themselves, measured in a large aperture (5" in diameter), was compared to measurements from the 2MASS point source catalog. The latter calibrations have a larger error, since most of the QSOs are barely detected in 2MASS. 

\section{Double Structure in the H-band Images}

\begin{figure*}[t]
\figurenum{1}
\label{sdss_qsos}
\centering
\includegraphics[width=\textwidth]{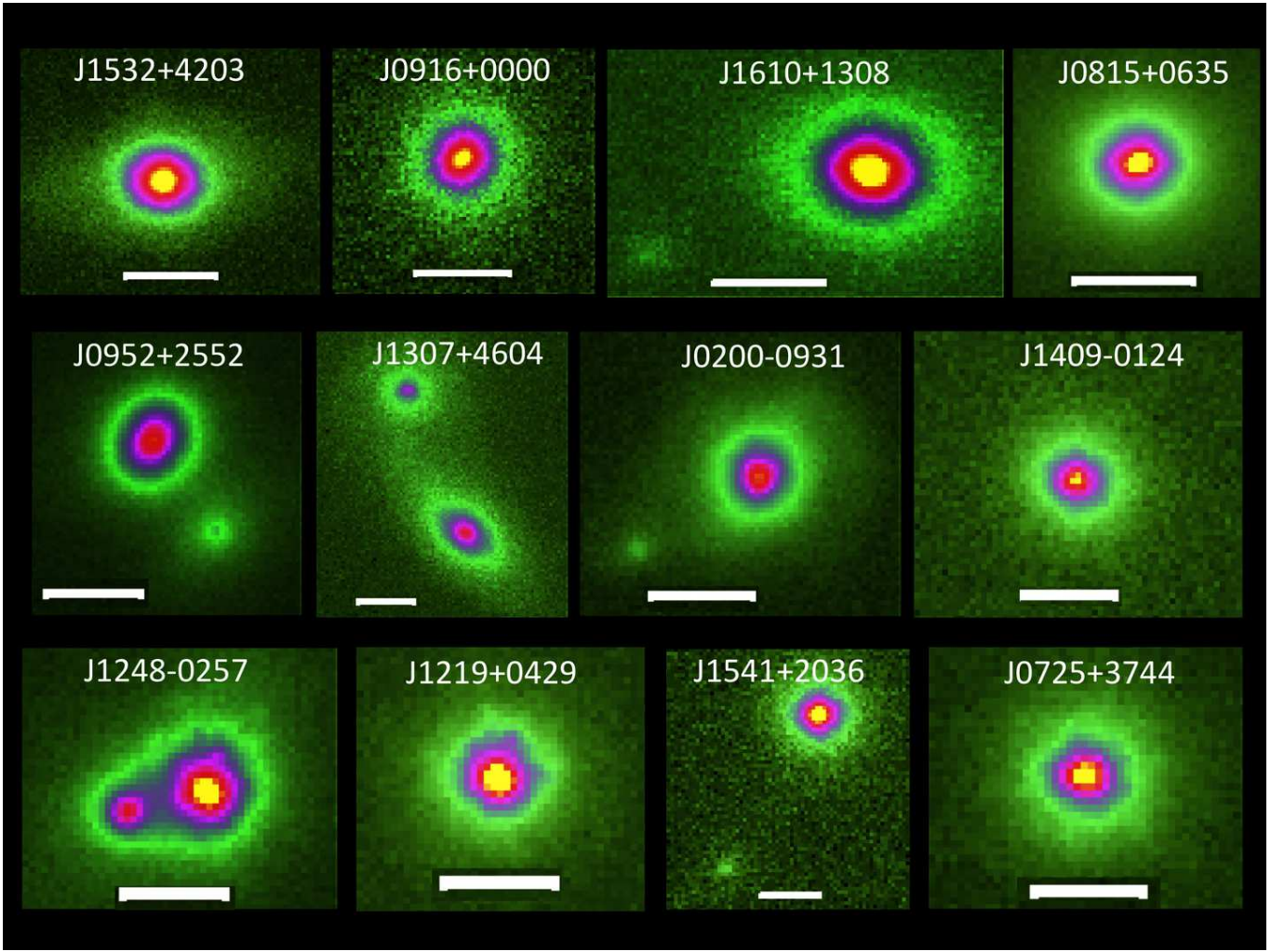}
\caption{ Keck laser guide star AO H-band images of double-peaked narrow line SDSS QSOs, taken
using the NIRC2 camera. The scale bar corresponds to 4 kpc at the redshift of the system. Six of the targets show signs of merger activity or close companions.
}
\end{figure*}

The final H-band images of the 12 QSOs are presented in Fig.~\ref{sdss_qsos}. Our images resolve structure on scales of a kpc or less in all these galaxies. A large fraction of our targets appear to have companions or show signs of being in a galaxy merger, highlighting the value of kinematic methods to find AGN in mergers. Of the twelve, three are in pairs with fairly equal H-band luminosity ratio: J095207+255257, J130724+460400 and J124859-025730. The obvious signatures of an interaction in these systems strongly suggest that they are undergoing a merger event. J020011-093126 has a minor companion, but a low surface brightness bridge connects the two galaxies. We will refer to these systems as `mergers' in the rest of the paper. Two more galaxies appear to have significantly fainter companions:  J161027+130806 and J154107+203608. These may be minor mergers or satellite systems, or the fainter object may be entirely unrelated to the AGN host. We call these `minor doubles'.

For the systems with double structure, the separation between the centers of the two galaxy components
are listed in Table 1, both in arcseconds and kpc. The four clear mergers have separations ranging from 12 kpc down to 3 kpc, and three of the four are separated by less than 6 kpc. Note that the angular diameter of the SDSS spectroscopic fiber aperture ($3"$) corresponds to distances of $5-15$ kpc over the redshift range of our galaxies. For some of the pairs with large separations ($> 10 $ kpc), it is not clear that sufficient nuclear light from the neighboring galaxy will have entered the SDSS fiber to produce double-peaked [\ion{O}{3}] lines, in which case the line splitting may be produced by processes related to just one of the nuclei. This is particularly relevant for J072554+374436, which has a compact point-like neighbor that is $2.9$'' away (not shown in Fig.~\ref{sdss_qsos}). This separation is about the diameter of the SDSS aperture. In this case, it is highly unlikely that the second component to the \othree\ line came from the nearby object, which, based on its compact appearance, may be a faint foreground star. In general, the chance of random alignment between one of the QSOs and a faint un-related galaxy in the foreground or background is quite small (a few percent), based on source counts of galaxies in the H-band from NIR surveys \citep[e.g.][]{frith06}. Therefore, most of the double galaxies in this NIRC2 imaging sample are expected to be real associations.

We modeled the H-band images of the AGNs using the galaxy structure fitting code \anchor{http:users.obs.carnegiescience.edu/peng/work/galfit/galfit.html}{GALFIT v3.0} \citep{peng02}. The brighter galaxy was modeled as a combination of a point source and an extended stellar light distribution having an elliptical Sersic profile with a variable Sersic index, half-light radius, ellipticity and orientation. In the cases with double structure, we modeled the fainter galaxy with a single elliptical Sersic profile, making the assumption that the point-like QSO is to be found only in the brighter object of the pair (as discussed in \S5.2). Each fit was examined by eye, and obvious failures were re-run with a different set of initial parameters until a good fit was achieved. The primary aim of these fits was to estimate the total H-band luminosity of the nuclear point source and the host galaxy, as well as the companion, if applicable. These estimates are listed in the last three columns of Table 1, where m$_{\textrm{H}}$ is the observed magnitude in the H-band in the 2MASS magnitude system. In a forthcoming paper, we will present a more detailed and accurate treatment of the structural analysis of our sample, including comparisons to NIRC2 images of inactive galaxies from the \anchor{http://irlab.astro.ucla.edu/cats/index.shtml}{Center for Adaptive Optics Treasury Survey} \citep{koo07}.

While these targets are all selected to be optical Type I AGN, it is clear that not all are dominated by a central point source. Extended host light is seen in most of these galaxies. This is partly due to the fact that we chose to image, when possible, objects that appeared to have a non-negligible fraction of their light coming from the host in the SDSS spectra. The AGN point source contributes from 25\% to 100\% of the total light in most of these systems, 
though in two cases, the estimated AGN fraction is much lower. The first, J130724.08+460400.9, is a pair of 
clearly interacting widely separated bright galaxies and the point source contributes less than 10\%\ of
the total light of the larger galaxy. The image of the second case, J091649.41+000031.5, shows an elliptical galaxy with a fairly smooth core and the GALFIT modeling suggests a negligible amount of light from a point source. 
However, our PSF monitoring images, taken just after the images of the galaxy, show
that the performance of the AO system suffered significant variation during this phase of the night. 
A broad PSF, probably due to an unsettled low bandwidth wavefront sensor in 
the AO system, would smooth out the core of the image and lower the contrast of a point source. 
Therefore, we are unable to accurately estimate the point source fraction in this galaxy, but can 
conclude that it does not dominate the H-band light. This example serves to demonstrate the importance of concurrent PSF monitoring in AO studies of the structure of distant galaxies. 

\section{Do Both Galaxies Host AGNs?}

\begin{figure}[t]
\figurenum{2}
\label{o3_hb_ratio}
\centering
\includegraphics[width=\columnwidth]{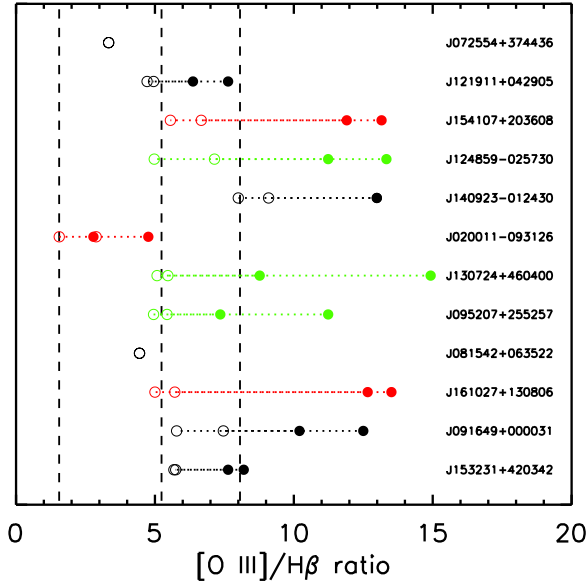}
\caption{A graphical comparison of \othree/H$\beta$ ratio of the two velocity components, measured from SDSS spectra, of the QSOs with NIRC2 imaging. Imaged objects with no double structure are shown with black points, while red is used for objects with large luminosity ratios between the galaxies in the image (minor doubles) and green is used for objects with roughly equal luminosity ratios (mergers). The best estimates for the ratio are shown as filled points, while the 2$\sigma$ lower limits to the ratio are plotted with open points; dotted lines join the measured values and limits, to guide the eye. Two objects have undetected narrow H$\beta$ and therefore only have lower limits on the ratio. The vertical dashed lines are three representative values of \othree/\hb\ along the curve separating star-forming galaxies from composite/AGN systems, as defined by \cite{kauffmann03}. The highest value ($\sim 8$) is for the lowest mass star-forming galaxies generally found in the SDSS, the middle value of $\sim 5$ is for typical low-mass star-forming galaxies and the lowest value corresponds to typical high mass galaxies. The majority of the velocity components have ratios consistent with AGN ionized gas.}
\end{figure}

We have shown that double galaxy structure is seen in a large fraction of QSOs which show double peaked narrow-lines. However, the question arises whether the fainter of the two galaxies also hosts an active nucleus, i.e, are these systems true double AGN? While sensitive to mergers, our NIR images do not provide color information about the galaxies or their point sources. From the single-band images alone, we are unable to test whether an AGN exists in each nucleus in the cases of the galaxies with double structure. However, we can carry out a simple test which compares the ionization level of both [\ion{O}{3}] velocity components seen in the SDSS spectra.  This allows us to determine if one of the peaks comes from, say, a low metallicity star-forming satellite galaxy. 


We concentrate on the ionization-sensitive \othree/\hb\ ratio which involves two lines that are close in wavelength and therefore are not strongly influenced by differential extinction or by the varying resolution of the SDSS spectrograph across its spectral range. This ratio was measured by first modeling the \othree\ line from each object as the sum of two independent gaussian profiles, which were then scaled in flux to match the narrow \hb\ line. In all cases, the broad-line \hb\ and the local continuum level were concurrently fit by eye using a high order polynomial with a variable index. Uncertainties in the fitting of the continuum were incorporated into the error in the normalization of the gaussian components.

In Fig.~\ref{o3_hb_ratio}, we show \othree/\hb\ for each of the two [\ion{O}{3}] lines in the twelve QSOs in our sample. From this figure, it is clear that most of our measured line ratios are above the highest possible star-forming value for normal galaxies, so these components are most likely AGN. Almost all lie to the right of the middle star-forming value, implying that AGN ionization dominates the narrow \hb\ line in most of these objects. 

The one case with low \othree/\hb\ ratios, J020011.52-093126.1, is surprising, since the image clearly shows a bright point source and the spectrum is dominated by strong QSO spectral features. There is no doubt that a bright AGN is present in this system. Interestingly, this is a double system showing some signatures of a merger. Both velocity peaks show a low ratio, indicating that the \hb\ line is enhanced relative to \othree , with respect to the typical AGN, for both components in this system. The low level of ionization in both peaks suggests that both components come from a common ionized region, supporting an outflow as a likely explanation for the line-splitting, rather than a merger. This implies that some QSOs that are in mergers may have a double [\ion{O}{3}] profile that is unrelated to the merger activity.

Five of the AGNs with double structure in our images lie to the right of the vertical line for which \othree/\hb\ $=8$, whereas only two of the non-double AGNs are in this region. This may point to a subtle trend that AGN in close double systems tend to have higher overall levels of nuclear ionization, possibly due to different nuclear gas environments mediated by the interaction. However, a significantly larger sample of isolated and merging AGN is needed to test this difference in a statistically meaningful way.

\section{Discussion}

\subsection{AGN luminosities}

\begin{figure}[t]
\figurenum{3}
\label{lumin}
\centering
\includegraphics[width=\columnwidth]{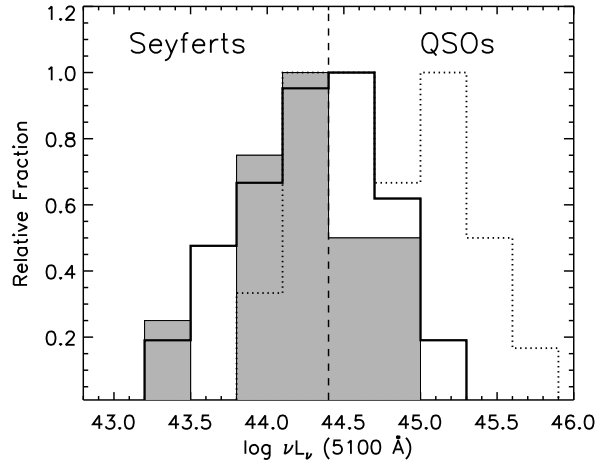}
\caption{Distributions of monochromatic luminosities at 5100 \AA. The open solid histogram is the distribution
for the parent sample of SDSS Type I AGN with double-peaked narrow lines, which may be compared
to the filled histogram, which is the distribution for the 12 AGNs with H-band imaging presented in this paper.
The dashed histogram is the distribution for 28 PG-QSOs from the QUEST survey. Our sample is equally divided
among high luminosity Seyferts and low luminosity QSOs.}
\end{figure}

In Fig. \ref{lumin}, we compare the total $5100$ \AA\ monochromatic luminosity distributions of the 12 AGNs in our sample
(filled histogram) with that of the parent sample of double-peaked Type I AGN from the SDSS (solid open histogram).
The traditional divide between Seyferts and QSOs (log $\nu$L$_{5100}$ = 44.37) is indicated by a vertical dashed line.
Our sample has a range of about an order of magnitude in luminosity and spans the divide between luminous Seyfert
1s and QSOs, with a median value of 44.32, slightly below the median of the parent sample. We also include in the
figure, as a dashed histogram, the distribution of luminosities for a sample of local PG-QSOs
from the Quasar/ULIRG Evolution Survey \citep[QUEST,][]{veilleux09}. Our objects overlap with the faint end of the
QUEST distribution, but we do not have any AGNs that sample the high luminosity end among local QSOs, which
can be more than an order of magnitude brighter than our brightest objects. One must keep in mind that
the following discussion is valid only for AGNs that lie at the transition between normal Seyferts and QSOs, 
and our results may not be applicable to the most luminous QSOs.

\subsection{Black Hole Masses and Growth}

\begin{figure*}[t]
\figurenum{4}
\label{bh_masses}
\centering
\includegraphics[width=\textwidth]{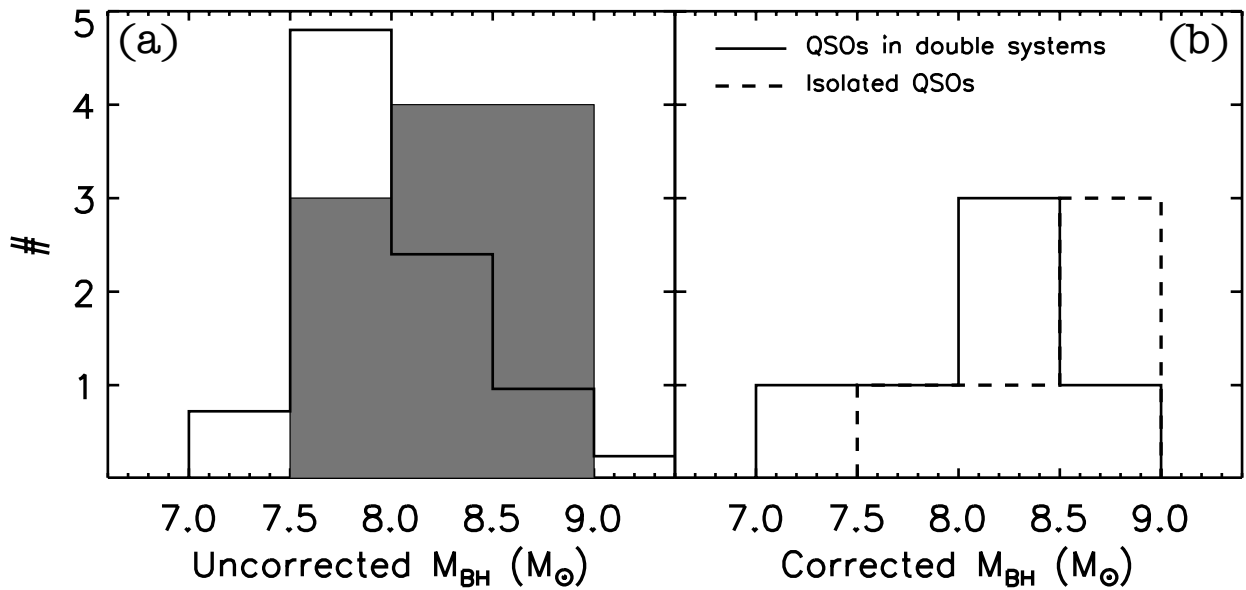}
\caption{Black Hole masses of the double-peaked [O III] QSO sample. The left panel (a) compares the distribution of $M_{BH}$ of the parent SDSS double-[O III] sample (open histogram, scaled down in number) with the masses of  all 12 QSOs in our sample (solid histogram), uncorrected for host galaxy contamination. The galaxies in our sample typically host more massive black holes than the average SDSS double-[O III] QSO. The right panel (b) compares $M_{BH}$ between AGN in double systems (solid line) and isolated systems (dashed line), corrected for the light of the host galaxy. While a difference in the distribution is seen, it is not significant, given the small number of objects in each sub-sample. }

\end{figure*}

With a set of genuine Type I AGN in pre-coalescence mergers, we can estimate SMBH masses relative to the masses of the merging hosts and may be able to test, with a sufficiently large sample, whether the black hole growth leads or lags the growth of stellar mass in the hosts. While our current sample of double AGN is too small to make many statistically significant statements, we nevertheless explore this possibility in order to understand the limitations and complexities of such an analysis. 

We start by making the weak assumption that each QSO system contain only one Broad-Line AGN. The second active nucleus, if present, is a Type II AGN, i.e, obscured along our line of sight so that only the narrow-line emission is visible.
Surveys of local AGN find that Type IIs outnumber Type Is by a factor of 3, possibly determined by the mean covering factor of the putative dust torus believed to exist in most AGN \citep{risaliti99}. However, there is some evidence that this difference may be smaller for distant or luminous AGN \citep{hasinger08}. Since our dual AGN are typically at the boundary between Seyferts and QSOs in terms of their nuclear bolometric luminosity, we will assume that Type II nuclei are more common and that all of the dual AGN in our sample are pairs of Type I and Type II nuclei. 

We follow the prescription of \citet[][Eqn. 2]{shields03} and estimate the masses of the SMBHs as follows:

\begin{equation}
M_{BH} \; = \; 10^{7.69} v_{3000}^{2} L_{44}^{0.5} \textrm{        solar masses}
\end{equation}

\noindent where $v_{3000}$ is the FWHM of broad \hb\ in units of 3000 \kms\, 
measured from the SDSS spectra and $L_{44}$ is the continuum luminosity of the QSO per decade in wavelength at 
a rest wavelength of $ 5100$ \AA, in units of $10^{44}$ erg s$^{-1}$. We follow the technique outlined in \citet{salviander07} for the measurement of the \hb\ line shape from SDSS spectra. After subtracting the narrow \hb\ emission line using a 1:10 scaled version of the \othree\ line, the remaining broad \hb\ line is modeled as a Gauss-Hermite function, from which $v_{3000}$ is derived. 

The continuum luminosity $L_{44}$ is also measured directly from the SDSS spectra of the QSOs. The light in the large aperture SDSS spectrum is a combination of emission from the QSO and from the host galaxies in the system. Therefore, we need to apply a correction factor to $L_{44}$ which accounts for host galaxy light within the SDSS spectral aperture. The correction term was estimated in the following way. First we calculated the flux ratio between the \emph{observed} H-band and the rest-frame $5100$ \AA\ (essentially a color: H$_{obs} - 5100$) from the Spectral Energy Distributions (SEDs) of an average SDSS QSO and a set of normal galaxies, taken from the publicly available SWIRE template library \citep{polletta07}. We calculate this color at the redshifts of each of the QSOs in our sample. We found that the difference in H$_{obs} - 5100$ for early-type and late-type galaxies is small at all our redshifts ($<0.5$ mag), so we adopted a typical value. 

We then smoothed our NIRC2 images to approximate ground-based seeing and, guided by our GALFIT fits to our NIRC2 images, measured the flux in the point source relative to the total light within a 3" circular aperture centered on the galaxy (or the brighter galaxy, in case of doubles). This measurement is equivalent to determining the point source fraction in an SDSS-like aperture, but in the observed H-band. Applying our estimates of H$_{obs} - 5100$,  we scaled the H-band point source fraction to get the point source fraction at rest-frame $5100$ \AA. Since QSO SEDs are much bluer than galaxy SEDs, the point-source fraction increases from a modest fraction of the light in the H-band (greater than 10\%) to 40\% to 100\% of the light at $5100$ \AA. Finally, we applied the point-source fraction at $5100$ \AA\ to our estimates of $L_{44}$ measured directly from the SDSS spectra, to get revised values of the QSO continuum luminosity. It transpired that, since M$_{BH}$ depends only on the square-root of $L_{44}$, the effect of this correction is only modest (less than 0.1 dex reduction in the median M$_{BH}$ of our sample). 

In Fig.\ref{bh_masses}a, the distributions of the uncorrected black hole masses (i.e, before applying the point-source fraction correction to $L_{44}$) of the QSOs in our sample (shaded histogram) are compared to the scaled distribution of the parent SDSS sample of \citet{smith10} (open histogram). We have excluded J091649.41+000031.5, as the point source fraction is very uncertain for this particular galaxy (see \S3). The median uncorrected BH mass of our AGN (log M$_{BH} = 8.43$) is significantly larger than that of the parent sample (log M$_{BH} = 7.94$). This difference is probably an effect of sample selection. Since we specifically targeted QSOs with some indication of host galaxy light in their SDSS spectra, we are likely to select systems with massive host galaxies and, by extension, massive black holes. Note that this same bias may come into play in studies that estimate QSO host galaxy velocity dispersions from SDSS spectra, such as in \citet{fu10}. Therefore, care must be taken in the interpretation of SMBH scaling relationships derived with these methods. 

In Fig.\ref{bh_masses}b, a comparison is made between the corrected M$_{BH}$ distribution (i.e, after accounting for the host galaxy) of the six QSOs which are in mergers or have companions against the corrected distribution of the five isolated QSOs (again excluding J091649.41+000031.5).  The median value of log M$_{BH}$ for the double systems is slightly lower than that of the isolated systems, by 0.14 dex, but the difference is not significant ( as determined by a two-sided Kolmogorov-Smirnov test). There is no statistical difference in this result if we restrict ourselves to only the four objects with major mergers. Therefore, we conclude that we do not detect any major differences in the $M_{BH}$ distribution between pre-merger and normal QSOs. If much SMBH growth occurs in QSOs, a substantial portion may not occur in the pre-merger phase. Major caveats do remain, however. If, for example, a large fraction of such systems harbor dual Type I AGN, in contrast to our in-going assumption in this analysis, then the $M_{BH}$ we calculated above are too high and will have to be adjusted. A larger sample of QSOs in mergers will help test this in the future. 

\subsection{The Incidence of Double AGN}

Taken at face value, the presence of 6 double or merging galaxies out of our sample of 12 QSOs implies that 
between 30\% - 70\% of Type I radio-undetected double-[\ion{O}{3}] emitters are in galaxy pairs at $0.2 < z < 0.6$, using binomial statistics of small numbers following \citet{gehrels86}. Of these, we find 4 (16\% - 56\%) in major mergers. These fractions compare favorably with those derived by a similar LGS AO program \citep{fu10} over the same redshift range. Given the small number of objects in our sample, we are unable to make any substantive statements about the evolution of this fraction, as \citet{fu10} have done, though they use a broader parent sample, including radio-detected and radio-loud AGN. We caution that the visibility of faint companions to the QSO hosts is strongly affected by surface brightness dimming towards higher redshifts and therefore a full treatment of the evolution of this fraction needs to take these factors into account.

We have verified, using H-band number counts from the literature, that chance alignments with background or foreground galaxies are not a likely explanation for the companions that we find around our QSOs. However, massive galaxies, such as those that host QSOs, typically reside in dense galactic environments. Therefore, the possibility exists that some or all of the minor doubles may be due to nearby satellite galaxies rather than ongoing mergers. 

To estimate the fraction of normal QSOs with faint neighbors, we examined a set of HST/NICMOS images of 28 QSO hosts published as part of the QUEST survey. The median redshift of the QUEST QSOs is lower than ours (about 0.14, compared to 0.35 for our NIRC2 sample). Despite this difference, we choose to use this comparative sample since it is derived from a homogenous high resolution H-band imaging dataset of normal QSOs with high quality and depth. Since the QUEST QSOs are not pre-selected to be double-peaked AGN and since the vast majority of QSOs do not show double-peaked narrow lines, we expect that the QUEST QSOs are predominantly normal QSOs that would not have made it into the \citet{smith10} catalog had they been at similar redshifts. Therefore, the neighbor statistics of these QSOs should be generally applicable to the SDSS QSO parent sample from which the \citet{smith10} sample was selected. We also assume that there has not been major evolution in the population of QSO host satellites between $z\sim 0.1$ and $z\sim 0.4$. 

From an examination of the QUEST images, we determine that approximately 35\%\ of 22 radio-quiet QSOs have a fainter companion within 15 kpc, but only two ($\sim 9$\%) show obvious spatially-resolved signs of merger activity. To compare, our double-peaked QSO sample has roughly the same fraction of minor doubles, but a larger fraction of mergers. Under the assumption that the statistics for the QUEST sample apply to the SDSS parent population of QSOs, we conclude that some or all of the minor doubles in our sample may arise from satellite galaxies around QSO hosts. If these satellites also host AGN, then the double-peaked narrow lines may indeed come from two separate active nuclei. However, it is also possible that the double-peaked structure in the minor doubles could come solely from the active nucleus containing the QSO, as in the 50\%\ of the NIRC2 sample that do not show double structure. At present, we are unable to differentiate between these two possible explanations for the minor doubles.

On the other hand, the rate of major mergers in our sample is considerably higher than those seen among QUEST QSOs, which implies that these mergers are likely to be the source of the double-peaked lines and are probably double AGN. In addition, the two cases of on-going massive mergers among the QUEST sample appear qualitatively different from the mergers in our sample, in that they seem to have double nuclei that are at closer separations and are probably in a more advanced merger stage. 

\subsection{ Fraction of QSOs with Dual SMBHs }

Assuming that each of the two \othree\ peaks corresponds to a galaxy in the double systems, we are able to calculate the fraction of QSOs that host dual active SMBHs on scales of few to several kpc. \citet{smith10} estimate that 0.9\% of radio-undetected QSOs show double-peaked narrow AGN lines. Applying the fraction that show resolved double structure in our NIRC2 images, about 1 of every 200 QSOs (0.5\%) contain double accreting SMBHs. If, instead, we assume that the minor doubles are not double AGN, this ratio drops to 0.32\%, or 1 in 300 QSOs, though this may be as high as 0.5\% at the
$1\sigma$ upper uncertainty in our estimate. We denote this fraction by $f_{dual}$. Note that we independently 
find the same fraction of mergers as \citet{fu10} among double-peaked AGN. Since \citet{fu10} present a 
larger sample of imaged objects, we expect that the total uncertainty in $f_{dual}$ should be smaller 
than our estimate above, but we will adopt the more conservative uncertainty from our limited sample. 

We can compare $f_{dual}$ to a rough estimate of the number of dual AGN expected among QSOs.   Given the assumption that all QSOs are associated with a major merger event, let $\Gamma$ be the `duty cycle' of QSO activity during a typical merger, i.e, it is the fraction of the time during a typical merger when either one of the SMBHs is active. Let $\Delta$ be the fraction of the time in which a typical merger is in its early stages, with nuclei separated by kpc distances or more. Then, the fraction of QSOs which have double nuclei that we may observe with our NIRC2 program is:

\begin{equation}
f_{dual} = \Gamma \Delta
\end{equation}

\noindent where we have assumed that the activity of the two SMBHs in a merger are independent of each other, and that the activity also does not depend on the stage of a merger. This last assumption is unlikely to be true - simulations suggest that most of the activity happens at the latter stages of a merger. We include it here for simplicity and use our final calculations to test its validity.  

Morphological studies of mergers at various stages of development, as well as numerical simulations \citep[e.g.,][]{callegari09}, imply that most merging galaxies spend a substantial fraction of the time of the merger in a pre-coalescence phase with widely separated nuclei. In other words, $\Delta$ should be a fairly large fraction. Given $f_{dual} \approx 0.003-0.0065$, Eq.~2 places $\Gamma$ in a similar range. How does this compare to simple estimates of the QSO merger `duty cycle' from population statistics?

If we assume the QSOs are found only among relatively massive galaxies ($ \log \textrm{M}_* > 10.5$ M$_{\odot}$), we can place some constraints on $\Gamma$ by comparing the merger rates of such galaxies with the frequency of QSOs among them.  Based on the analysis of morphologies and pair statistics from field galaxy surveys, roughly 4\%\ of massive ($\sim \textrm{M}^{*}$) galaxies host on-going galaxy mergers \citep{bundy09, hopkins10}. This fraction has some uncertainty, but is unlikely to be greater than 10\%. From the ratio of the space densities of  $10^{11}$ M$_{\odot}$ galaxies \citep[$\sim 10^{-3.5}$ Mpc$^{-3}$ dex$^{-1}$,][]{bundy06} and QSOs in the luminosity range of our sample \citep[$-23< M_{V} < -21$, $10^{-6}$ Mpc$^{-3}$ mag$^{-1}$,][]{croom04}, at $z\sim 0.5$, QSOs are found in only about 0.3\%\ of massive galaxies. This fraction is roughly consistent with the more detailed analysis by \citet{hamilton02} at $z=0.26$. The QSO `duty cycle' $\Gamma$ is given by the ratio of the fraction of QSOs among galaxies to the fraction of mergers among galaxies, as long as all QSOs are associated with mergers. Therefore, $\Gamma \sim 0.003/0.04 \approx 0.08$. This estimate is significantly different than the value we derived earlier from dual AGN statistics. Put simply, the fraction of time a merging pair of galaxies spends in a QSO phase is more than an order of magnitude higher than an estimate based on our measurement of $f_{dual}$. 

There are several possible explanations for this discrepancy and we briefly highlight a few here.
The simplest is that our estimate of the value of $f_{dual}$, with its considerable uncertainty, is not representative of the true population of AGN in this luminosity range. However, our fractions match well with a larger sample from \citet{fu10} and are unlikely to be incorrect by an order of magnitude (i.e, we measure only 30\% in mergers, when the true fraction is almost 100\%).

An explanation that is outlined in \citet{smith10} suggests that searches for double-peaked emitters are highly incomplete and miss as many as  90\%\ of true mergers, due to line of line-of-sight effects. This would increase $f_{dual}$ and bring into closer agreement the two estimates of $\Gamma$. Alternatively, a majority  of SDSS Type I AGN may not be associated with massive mergers, but driven by processes like minor mergers or secular infall, as recently proposed for X-ray selected Seyfert galaxies with similar AGN luminosities at intermediate redshifts \citep{cisternas10}. High resolution studies of QSO host structure \citep{guyon06} find that significant fractions ($\sim 40$\%) have predominantly disk morphologies, inconsistent with major mergers fueling these systems. In this case, the estimate of $\Gamma$ from population statistics is not a true duty cycle, but is effectively diluted by the population of non-merger QSOs. A third possibility is that unobscured activity may be substantially suppressed in the early stages of a merger. In such a scenario, merging SMBHs separated by more than a few kpc are much less likely to achieve high accretion rates and exhibit QSO luminosities than SMBHs close to or after final coalescence. This notion jibes with the results of merger simulations \citep[e.g.,][]{hopkins08}. At present, given the limited statistics, we cannot easily distinguish which of these reasons is most significant in explaining the difference between the dual AGN fraction we measure and that expected from the statistics of the QSO population. All three may play a role to some degree. In addition, the dual AGN fractions we calculate here are applicable most directly only to moderate luminosity Type I AGNs. Dual fractions among luminous QSOs may be higher and systematic spectroscopic and imaging programs of luminous QSOs at high spatial resolution could throw some light on the fueling properties of strongly accreting AGN.




\section{Conclusions and Future work}

From a well-defined sample of 12 SDSS radio-quiet Type I QSOs that display double-peaked structure in the \othree\ line, we identify double structure in 6 objects, using NIRC2 LGS AO imaging. We verify that both [O III] velocity components in the QSOs are likely to come from AGN-ionized gas, consistent with the hypothesis that AGN exist in both galaxies in the six double systems. We model the images with GALFIT to estimate galaxy and point source luminosities, from which we estimate SMBH masses. There is no significant difference in $M_{BH}$ between isolated and merging QSOs in our imaging sample, though the mean mass of our sample is higher than the parent SDSS population. Finally, we estimate that mergers detectable by current Keck LGS AO and NIRC2 imaging exist in 0.3\%-0.65\%\ of QSOs, which is considerably lower than the merger rate among inactive massive galaxies. 

In addition to developing improved morphology statistics via a larger imaging sample, an attractive next step would be to use AO spatially resolved IFU spectroscopy of our smallest-separation targets to answer the key question of whether each galaxy in our pairs contains an active nucleus. We are currently pursuing this avenue are part of a follow-up study of the best dual AGN candidates.

\acknowledgments

Data presented herein were obtained at the W. M. Keck Observatory,
which is operated as a scientific partnership among the
California Institute of Technology, the University of California,
and the National Aeronautics and Space Administration. The
Observatory and the Keck II AO system were both made possible
by the generous financial support of the W. M. Keck Foundation.
The authors wish to extend special thanks to those of
Hawaiian ancestry, on whose sacred mountain we are privileged
to be guests. Without their generous hospitality, the observations
presented herein would not have been possible.

This material is based in part upon work supported by the National Science Foundation under award number AST-0908796. McGurk is supported by a Graduate Research Fellowship from the National Science Foundation.

Funding for the SDSS and SDSS-II has been provided by the Alfred P. Sloan Foundation, the Participating Institutions, the National Science Foundation, the U.S. Department of Energy, the National Aeronautics and Space Administration, the Japanese Monbukagakusho, the Max Planck Society, and the Higher Education Funding Council for England. The SDSS Web Site is http://www.sdss.org/.


\begin{thebibliography}{dummy}

\bibitem[Bundy et al.(2006)]{bundy06}
Bundy, K., Ellis, R.S., Conselice, C.J., Taylor, J.E., Cooper, M.C., Willmer, C.N.A., Weiner, B.J., Coil, A.L., Noeske, K.G., \& Eisenhardt, P.R.M. 2006, \apj, 651, 120

\bibitem[Bundy et al.(2009)]{bundy09}
Bundy, K., Fukugita, M., Ellis, R.S., Targett, T.A., Belli, S., \& Kodama, T. 2009, \apj, 697, 1369

\bibitem[Callegari et al.(2009)]{callegari09}
Callegari, S., Mayer, L., Kazantzidis, S., Colpi, M., Governato, F., Quinn, T., \& Wadsley, J. 2009, \apjl, 696, 89

\bibitem[Comerford et al.(2009)]{comerford09}
Comerford, J., et al. 2009, \apj, 698, 956

\bibitem[Cisternas et al.(2010)]{cisternas10}
Cisternas, M., et al. 2010, arXiv:1009.3265

\bibitem[Croom et al.(2004)]{croom04}
Croom, S. M., Smith, R. J., Boyle, B. J., Shanks, T., Miller, L., Outram, P. J., \& Loaring, N. S. 2004, \mnras, 349, 1397

\bibitem[Ferrarese \& Merritt(2000)]{ferrarese00}
Ferrarese, L., \& Merritt, D. 2000, \apj, 539, L9

\bibitem[Foreman et al.(2009)]{foreman09}
Foreman, G., Volonteri, M., \& Dotti, M. 2009, \apj, 693, 1554

\bibitem[Frith et al.(2006)]{frith06}
Frith, W. J., Metcalfe, N., \& Shanks, T. 2006, \mnras, 371, 1601

\bibitem[Fu et al.(2010)]{fu10}
Fu, H., Myers, A.D., Djorgovski, S.G., \& Yan, L. 2010, arXiv:1009.0767

\bibitem[Gebhardt et al.(2000)]{gebhardt00}
Gebhardt, K., et al. 2000,\apj, 539, L13

\bibitem[Gehrels(1986)]{gehrels86}
Gehrels, N. 1986, \apj, 303, 336

\bibitem[Guyon et al.(2006)]{guyon06}
Guyon, O., Sanders, D.B., \& Stockton, A. 2006, \apjs, 166, 89

\bibitem[Hamilton et al.(2002)]{hamilton02}
Hamilton, T. S., Casertano, S., \& Turnshek, D. A. 2002, \apj, 576, 61

\bibitem[Hasinger(2008)]{hasinger08}
Hasinger, G. 2008, \aap, 490, 905

\bibitem[Hennawi et al.(2006)]{hennawi06}
Hennawi, J. et al. 2006, \aj, 131, 1

\bibitem[Hennawi et al.(2010)]{hennawi10}
Hennawi, J. et al. 2010, \apj, 719, 1672

\bibitem[Hopkins et al.(2006)]{hopkins06}
Hopkins, P., Hernquist, L., Cox, T.J., Di Matteo, T., Robertson, B., \& Springel, V., 2006, \apjs, 163, 1

\bibitem[Hopkins et al.(2008)]{hopkins08}
Hopkins, P., Hernquist, L., Cox, T.J.,  \& Kere\v{s}, D., 2008, \apjs, 175, 356

\bibitem[Hopkins \& Hernquist(2009)]{hopkins09}
Hopkins, P., \& Hernquist, L. 2009, \apj, 694, 599.

\bibitem[Hopkins et al.(2010)]{hopkins10}
Hopkins, P.F., Bundy, K., Croton, D., Hernquist, L., Kere\v{s}, D., Khochfar, S., Stewart, K., Wetzel, A., \& Younger, J.D. 2010, \apj, 715, 202

\bibitem[Junkkarinen et al.(2001)]{junkkarinen01}
Junkkarinen, V., Shields, G. A., Beaver, E. A., Burbidge, E. M., Cohen, R. D., Hamann, F., \& Lyons, R. W. 2001, \apj, 549, L155

\bibitem[Kauffmann et al.(2003)]{kauffmann03}
Kauffmann, G., et al 2003, \mnras, 346, 1055 

\bibitem[Koo et al. (2007)]{koo07}
Koo, D.C., Melbourne, J., Max, C., Metevier, A., Ammons, S.M., Larkin, J.E., Barczys, M., Wright, S.A., \& Steinbring, E. 2007, Proc. IAU Symposium 235, CUP, 355

\bibitem[Kuo et al.(2008)]{kuo08}
Kuo, C.-Y., Lim, J., Tang, Y.-W., \& Ho, P. T. P. 2008, \apj, 679, 1047

\bibitem[Liu et al.(2009)]{liu09}
Liu, X., Shen, Y., Strauss, M.A., \& Greene, J.E. 2010, \apjl, 708, 427

\bibitem[Myers et al.(2007)]{myers07}
Myers, A.D. et al., 2008, \apj, 658, 99

\bibitem[Myers et al.(2008)]{myers08}
Myers, A.D. et al., 2007, \apj, 678, 635

\bibitem[Peng et al.(2002)]{peng02}
Peng, C.Y. et al., 2002, \aj, 12, 266

\bibitem[Polletta et al.(2007)]{polletta07}
Polletta, M., et al. 2007, \apj, 663, 81

\bibitem[Risaliti et al.(1999)]{risaliti99}
Risaliti, G., Maiolino, R., Salvati, M. 1999, \apj, 522, 157

\bibitem[Rosario et al.(2010)]{rosario10}
Rosario, D.J., Shields, G.A., Taylor, G.B., Salviander, S., \& Smith, K.L. 2010, \apj, 716, 131

\bibitem[Salviander et al.(2007)]{salviander07}
Salviander, S., Shields, G.A., Gebhardt, K., \& Bonning, E.W. 2007, \apj, 662, 131

\bibitem[Shields et al.(2003)]{shields03}
Shields, G.A., Gebhardt, K., Salviander, S., Wills, B.J., Xie, B., Brotherton, M.S., Yuan, J., \& Dietrich4, M., 2003, \apj, 583, 124

\bibitem[Smith et al.(2010)]{smith10}
Smith, K.L., Shields, G.A., Bonning, E. W., McMullen, C. C., Rosario, D. J., \& Salviander, S. 2010, \apj, 716, 866

\bibitem[Veilleux et al.(2009)]{veilleux09}
Veilleux, S., et al. 2009, \apj, 701, 587

\bibitem[Wang et al.(2009)]{wang09}
Wang, J., Chen, Y, Hu, C., Mao, W., Zhang, S., Bian, W. 2009, \apjl, 705, 76

\bibitem[Whittle(1992)]{whittle92}
Whittle, M. 1992, \apj, 387, 121

\bibitem[Wilson \& Willis(1980)]{wilson80}
Wilson, A.S., \& Willis, A.G. 1980, \apj, 240, 429

\bibitem[Wizinowich et al.(2006)]{wizinowich06}
Wizinowich, P.L. et al.2006, \pasp, 118, 297

\end{thebibliography}
\end{document}